\documentclass[14pt,preprint2,longabstract]{aastex}
\newcounter{column_number}
\shorttitle{A {\it Chandra} survey of Galactic globular clusters}
\shortauthors{Cheng et al.}
\begin{document}
\title{Exploring the Mass Segregation Effect of X-ray Sources in Globular Clusters. II. The Case of Terzan 5}
\author{Zhongqun Cheng$^{1,2,3}$, Zhiyuan Li$^{2,3}$, Taotao Fang$^{1}$, Xiangdong Li$^{2,3}$ and Xiaojie Xu$^{2,3}$}
\affil{$^{1}$ Department of Astronomy and Institute for Theoretical Physics and Astrophysics, Xiamen University, Xiamen, Fujian 361005, China} 
\affil{$^{2}$ School of Astronomy and Space Science, Nanjing University, Nanjing 210023, China} 
\affil{$^{3}$ Key Laboratory of Modern Astronomy and Astrophysics (Nanjing University), Ministry of Education, Nanjing 210023, China}
\email{zqcheng@xmu.edu.cn; fangt@xmu.edu.cn}

\begin{abstract}
Using archival {\it Chandra} observations with a total effective exposure of 734 ks, we derive an updated catalog of point sources in the massive globular cluster Terzan 5. 
Our catalog covers an area of $58.1\, \rm arcmin^{2}$ ($R\leq 4.3 \, \rm arcmin$) with 489 X-ray sources, and more than $75\%$ of these sources are first detected in this cluster.
We find significant dips in the radial distribution profiles of X-ray sources in Terzan 5, with the projected distance and width of the distribution dips for bright ($L_{X} \gtrsim 9.5\times 10^{30} {\rm\ erg\ \,s^{-1}}$) X-ray sources are larger than that of the faint ($L_{X} \lesssim 9.5\times 10^{30} {\rm\ erg\ \,s^{-1}}$) sources. 
By fitting the radial distribution of the X-ray sources with a``generalized King model", we estimated an average mass of $1.48\pm0.11\,M_{\odot}$ and $1.27\pm0.13\,M_{\odot}$ for the bright and faint X-ray sources, respectively.  
These results are in agreement with that observed in 47 Tuc, which may suggest a universal mass segregation effect for X-ray sources in GCs.
Compared with 47 Tuc, we show that the two-body relaxation timescale of Terzan 5 is much smaller, but its dynamical age is significantly younger than 47 Tuc. These features suggest that the evolution of Terzan 5 is not purely driven by two-body relaxation, and tidal stripping effect also plays an important role in accelerating the dynamical evolution of this cluster.

\end{abstract}


\section{Introduction}
It has been recognized for decades that binaries play a pivotal role in the dynamical evolution of globular clusters (GCs), as they may serve as an important internal energy source in GCs (see \citet{hut1992a} for an early review). 
The binding energy stored in binaries in GCs is at least comparable to, and in many cases may well exceed, the total energy (in the form of the kinetic and potential energy of the single stars and the center-of-mass of the binaries) of the cluster as a whole \citep{heggie2003}. Through binary--single (b--s) and binary--binary (b--b) encounters, the binding energy locked up in close binaries can be extracted within the dense core of GCs, which is thought to be an effective ``heating mechanism" that supports the cluster against gravothermal collapse \citep{hut1983a,heggie2003}. The term ``binary burning" is used in analogy to hydrogen burning for stars. In the same way that hydrogen burning allows a star to remain in thermal equilibrium on the main sequence (MS) for a time much longer than the Kelvin--Helmholtz timescale, primordial binary burning allows a cluster to maintain itself in quasi--thermal equilibrium and avoid core collapse for a time much longer than the two-body relaxation timescale \citep{fregeau2003}.

For the numerous primordial binaries in GCs, simulations suggest that they may experience a variety of physical processes, which are absent in stellar evolution in isolation \citep{gao1991,hut1992b,sigurdsson1993,fregeau2004,chatterjee2010,morscher2013}. 
Under the effect of mass segregation, primordial binaries tend to sink to the dense core of GCs, where dynamical processes, mainly b--s and b--b encounters, would take place frequently. Depending on the binary hardness (defined as $\eta=|E_{b}|/E_{k}$, with $|E_{b}|$ the bound energy of binaries and $E_{k}$ the average kinetic energy of GC stars), dynamical encounters tend to transform hard binaries ($\eta>>1$) into harder systems (i.e., the binary burning process), whereas they turn soft MS binaries ($\eta<<1$) into softer systems or `ionize' them into single stars \citep{hills1975,heggie1975,hut1993}. 
It is also possible for the MS binaries to exchange one of their primordial members with other stars and lead to the formation of exotic binaries in GCs \citep{hills1976,rasio2000,ivanova2006,ivanova2008,ivanova2010,mapelli2014,hong2017}. Besides, close binaries may receive a large recoil velocity and escape from the cluster, and, alternatively, coalescence through gravitational radiation losses or collisional mergers \citep{gao1991,hut1992b,sigurdsson1993,fregeau2004,chatterjee2010,morscher2013,morscher2015,askar2017}.

Observationally, the identification of MS binaries in GCs is a tough work, mainly due to the low intrinsic luminosities of the low-mass stars and the crowded stellar environment. Fortunately, many exotic objects have been observed in GCs, which include blue straggler stars (BSS; \citealp{sandage1953}), low-mass X-ray binaries (LMXBs; \citealp{clark1975,katz1975}), millisecond pulsars (MSPs; \citealp{camilo2005,ransom2008}), cataclysmic variables (CVs) and coronally active binaries (ABs; \citealp{grindlay2001,pooley2002,edmonds2003a,edmonds2003b,heinke2005}). 
These objects are either very close binary systems (i.e., LMXBs, CVs, ABs) or the immediate remnants of close binaries (i.e., BSS, MSPs), and they may represent the final binary burning products of primordial binaries in GCs. Compared with normal stars, these exotic objects are found to be much more luminous (either in optical, X-ray or in radio band) and can be easily detected and picked out from the dense core of GCs, making them ideal tracers for studying stellar dynamical interactions and cluster evolution.

For example, the abundances (i.e., number per unit stellar mass) of LMXBs and MSPs are found to be orders of magnitude higher in GCs than in the Galactic field \citep{clark1975,katz1975,camilo2005,ransom2008}, which suggests that stellar dynamical interactions are in favour of creating these systems in GCs \citep{verbunt1987,pooley2003}. However, for CVs and ABs, because their progenitor MS binaries are mainly low-mass stars, they will evolve on a timescale comparable to or even greater than the GC relaxation time. \cite{davies1997} was among the first to predict that the formation of CVs through primordial channel should be suppressed in GCs, since a large fraction of CV primordial binaries would be dynamically disrupted in the dense cores of GCs. Cheng and colleagues, have confirmed the prediction of \cite{davies1997}, by finding a lower X-ray emissivity (thus a lower abundance of CVs and ABs) in most GCs with respect to the Galactic field \citep{cheng2018a}. In addition, they also found a strong correlation between binary hardness ratio (defined as the abundance ratio of X-ray-emitting close binaries to MS binaries) and cluster central stellar velocity dispersion in 30 GCs, which is consistent with the Hills-Heggie law of binary dynamical interactions in dense stellar systems \citep{cheng2018b}.
  
On the other hand, owing to the much lower mass in the progenitor MS binaries, \citet{davies1997} and \citet{belloni2019} argued that some primordial binaries might not have time to sink into the cluster center, instead they will evolve into CVs in the outskirts of GCs. Utilizing archival {\it Chandra} observations, \citet{cheng2019} performed a deep survey of weak X-ray sources in 47 Tucanae. They detected 537 point sources within a radius of $7.5\arcmin$ (corresponding to $8.77 \, \rm pc$) in 47 Tuc, and confirmed that there are some cluster X-ray sources located outside the cluster half-light radius ($3.7\, \rm pc$). More importantly, they showed that the radial distribution of X-ray sources in 47 Tuc is bimodal, with the observed surface density profile exhibiting a central peak, a dip at intermediate radii and a turnover at larger radii. 
These features are similar to that found in the radial distribution of BSS, and both are thought to be modulated by the mass segregation effect of primordial binaries in GCs \citep{ferraro2012,cheng2019}.

Terzan 5 is a heavily obscured (with an average color excess of $E(B-V) = 2.38$; \citealp{massari2012}) cluster located in the inner Galactic region ($l=3.8395^{\circ}$,$b=+1.6868^{\circ}$), at a distance of $d=5.9\pm0.5 \, \rm kpc$ from the Sun \citep{valenti2007}. 
Although generally considered as a GC \citep{harris1996}, the very nature of Terzan 5 is intricate, and at least two distinct populations of stars have been detected in Terzan 5, with star formation episodes separated by $\sim 7.5 \,\rm Gyr$ and stellar chemical abundance patterns are strikingly similar to those observed in the bulge field stars \citep{ferraro2009,ferraro2016}. 
These features, together with its huge mass ($M_c\sim 2\times 10^{6}M_{\odot}$; \citealp{lanzoni2010}), short and highly eccentric orbit (i.e., with perigalactic distance of $R_{p}=0.82\pm0.32\, \rm kpc$ and apogalactic distance of $R_{a}=2.83\pm 0.54\, \rm kpc$, respectively; \citealp{baumgardt2019}), suggesting that Terzan 5 could be a fossil remnant of the massive clumps that assembled the Galactic bulge \citep{ferraro2016}.
Among all Galactic GCs, Terzan 5 is outstanding as a prolific MSP factory, currently known to host a total of 37 MSPs \citep{cadelano2018}, which represent a quarter of the total population of pulsars in GCs\footnote{http://www.naic.edu/~pfreire/GCpsr.html}.
Indeed, Terzan 5 was found to have the highest stellar encounter rate among Galactic GCs \citep{bahramian2013}, making it one of the ideal laboratories for studying stellar dynamical interactions and exotic objects.

In this work, we present a study of mass segregation effect in Terzan 5, based mainly on the spatial distribution of X-ray sources.    
This paper is organized as follows. In Section 2 we present the {\it Chandra} observations and data preparation. 
In Section 3, we describe our source detection procedure and the resulting source catalog. 
We analyze the X-ray source radial distribution in Section 4, and explore its relation with GC mass segregation effect and Galactic tidal stripping effect in Section 5. 
A brief summary of our main conclusions is provided in Section 6.
  
\section{{\it Chandra} Observations and Data Preparation}

\begin{figure*}[htp]
\centering
\includegraphics[angle=0,origin=br,height=0.58\textheight, width=0.81\textwidth]{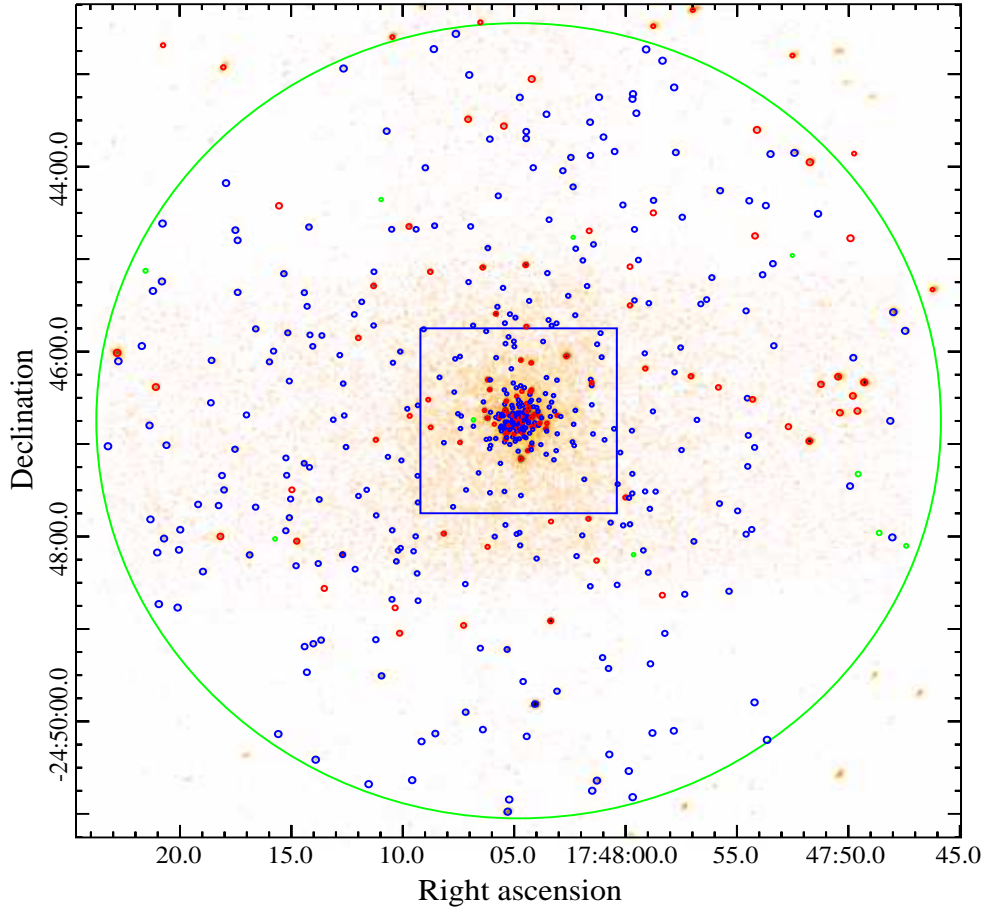}
\linespread{0.7}
\caption{The full-band (0.5-7 keV) {\it Chandra} merged image of Terzan 5. The images are smoothed with a Gaussian kernel with a radius of 3 pixels. Only X-ray sources located within the green circle ($R=4.3\arcmin$) have been detected and analyzed in this work. X-ray sources shown in blue are new detections of this work, while those in red and green were already identified by \citep{heinke2006}. We reidentified those sources in red but did not find those in green. A zoom-in view of the central $2\arcmin\times 2\arcmin$ region (blue square region) is illustrated in Figure-2. \label{fig:smoothimage}}
\end{figure*}
The inner region of Terzan 5 has been frequently visited by {\it Chandra}, primarily with its Advanced CCD Imaging Spectrometer (ACIS). A total of 24 ACIS observations were taken between July 2000 to July 2016. However, Terzan 5 was found to harbour three luminous transients, namely, EXO 1745-248, IGR J17480-2446 and Swift J174805.3-244637 \citep{bahramian2014}. When these sources were caught in outburst, in 6 observations (ObsID: 654, 655, 11051, 12454, 13161 and 13708), even just the PSF-scattered halo of the transients would severely affect our analysis of weak X-ray sources in Terzan 5. Hence, we chose to exclude these observations from our analysis. As listed in Table-1, all of the remaining 18 observations were taken with the aim-point at the S3 chip, among which 6 observations were performed with a subarry mode, to minimize the effect of pileup for bright sources.

We downloaded and uniformly reprocessed the archival data with CIAO v4.8 and CALDB v4.73, following the standard procedure\footnote{http://cxc.harvard.edu/ciao}. Briefly, we used the CIAO tool
{\it acis\_process\_events} to reprocess the level 1 event files, applying both charge transfer inefficiency and gain corrections. All of the observations were taken in the FAINT mode. To search for potential periods of high background, we created background light curves for each observation by excluding the source regions. No significant episodes of high background rates are observed for all but one observation (ObsID:3798), we therefore removed the episodes of background flares from this observation. The total effective exposure amounts to 734 ks at the center of Teran 5. 
To match and align the event images, we searched X-ray sources within the field of view (FoV) of each observation, and utilized the ACIS Extract (AE; \citealp{broos2010}) package to refine the source positions. The CIAO tool {\it reproject\_aspect} and {\it reproject\_event} were employed to calibrate the relative astrometry and projected the event files separately. ObsID 15615, which has the longest exposure, was served as the reference frame. We then merged the 18 observations into master event files using CIAO tool {\it merge\_obs}. three groups of images have been created in soft (0.5-2 keV), hard (2-7 keV) and full (0.5-7 keV) energy bands and with bin size of 0.5, 1 and 2, respectively. We also created exposure maps and exposure weighted averaged PSF maps for each of the three groups of files, following the procedures presented in \citet{cheng2019}. 
Our data preparation details are summarized in Table-1.

\section{X-ray Source Catalog}

\subsection{Source Detection and Sensitivity}
Following the procedures employed in a similar study for 47 Tuc \citep{cheng2019}, we employed a two stage approach to create the X-ray sources catalog in Terzan 5. Briefly, we first generate a liberal catalog of candidate point sources with {\it wavdetect} and visual examination. The {\it wavdetect} was run with aggressive thresholds to find weak sources, which are expected to enrol some spurious sources. We then extract and filter the candidate source list with AE, which allow us to evaluate the significance of weak sources interactively and iteratively\footnote{See Figure 8 of the AE User's Guide for details}. As suggested by the AE authors, such an interactive and iterative pruning strategy is helpful to identify weak sources in crowded environment \citep{broos2010}, since X-ray sources are overcrowded in the core of GCs, and extraction and pruning of point sources are inevitably affected by their surrounding neighbors. 

\begin{figure*}[htp]
\centering
\includegraphics[angle=0,origin=br,height=0.65\textheight, width=0.90\textwidth]{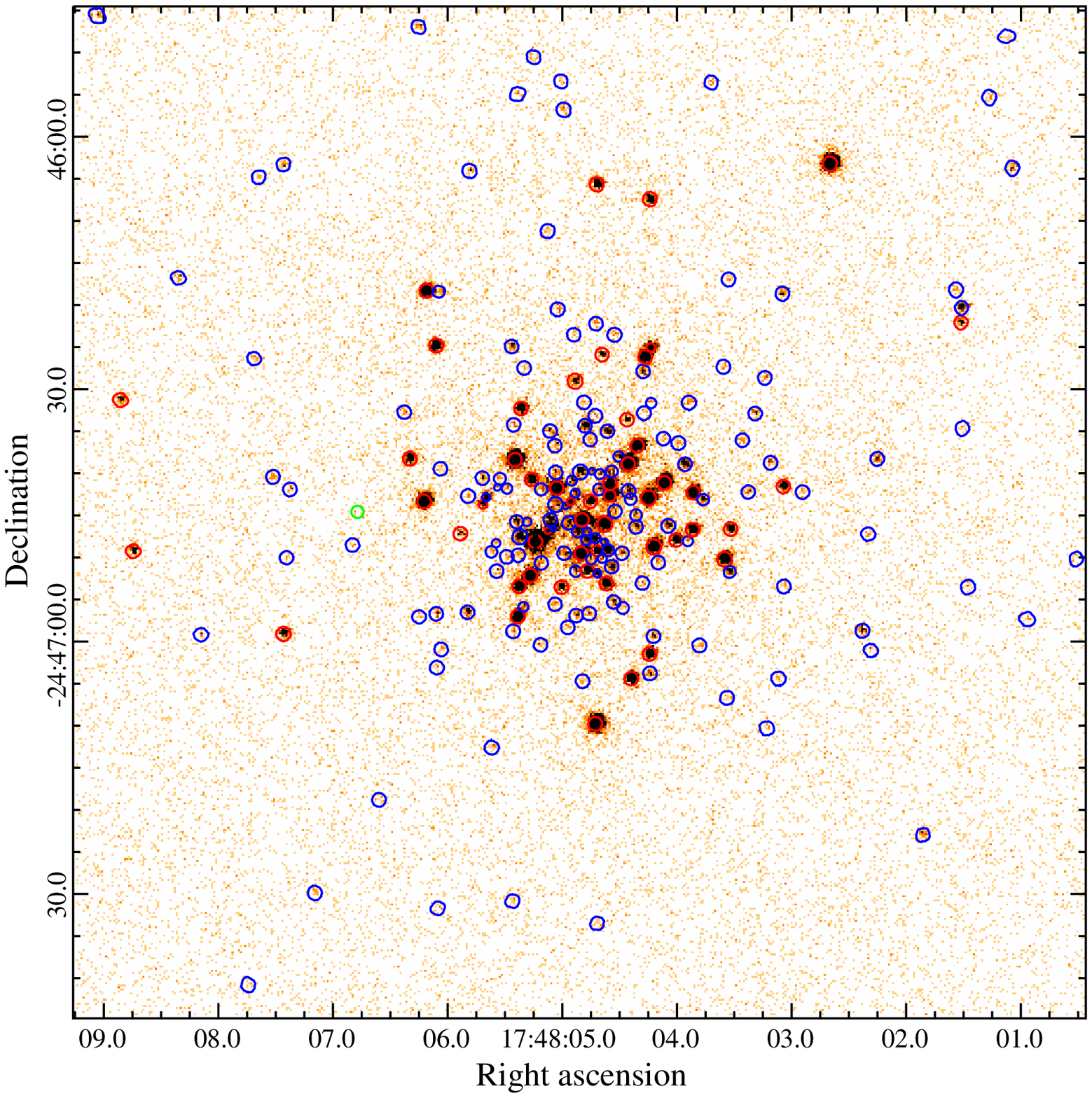}
\linespread{0.7}
\caption{{\it Chandra} merged image of the central $2\arcmin \times 2\arcmin$ region of Terzan 5. The image was rebinned to $0\arcsec.25$ $\rm pixel^{-1}$. 
The color-coded symbols denote the different types of sources, as in Figure-\ref{fig:smoothimage}. \label{fig:roominimage}}
\end{figure*}

To generate a candidate X-ray source list for Terzan 5, we ran {\it wavdetect} on each of the 9 combined images, using a ``$\sqrt{2}$ sequence" of wavelet scales (i.e., 1, 1.414, 2, 2.828, 4, 5.656 and 8 pixels). We confined the X-ray sources searching region as a circle with a radius of $4.3\arcmin$ (green circle in Figure~\ref{fig:smoothimage}) and adopted a false-positive probability threshold
of $1\times 10^{-6}$, $1\times 10^{-5}$ and $5\times 10^{-5}$ for event files with bin size of 0.5, 1.0 and 2.0 pixels, respectively. We also created an exposure time weighted averaged PSF map for each of the 9 merged images and supplied them to the {\it wavdetect} script. The {\it wavdetect} results were then combined into a master source list with the {\it match\_xy} tool from the Tools for ACIS Review and Analysis (TARA) packages\footnote{http://www2.astro.psu.edu/xray/docs/TARA/}. 
As experienced in the case of 47 Tuc, although the relatively loose source-detection threshold introduces a non-negligible number of spurious sources to the candidate list, a handful of faint sources are still missed by {\it wavdetect} within the crowded cluster core. Therefore, we picked out these sources visually and added them into the master source list, which resulted in a candidate list of 557 sources.

We then utilized AE to extract and examine the validation of each candidate source. Our source pruning processes are similar to the steps outlined in the validation procedure\footnote{http://www2.astro.psu.edu/xray/docs/TARA/ae\_users\_guide/procedures/} presented by the AE authors: All candidate sources have been extracted, merging with multiple extractions, pruning insignificant sources and repositioning repeatedly, until all the remaining sources are deemed significant.
AE provides an important output parameter (i.e., the binomial no-source probability $P_{B}$) to evaluate the significance of a source, which is defined as the probability of still observing the same number of counts or more assuming that the observed source is due to background fluctuations \citep{weisskopf2007}:
\begin{equation}
P_{\rm B}(X\ge S)=\sum_{X=S}^{N}\frac{N!}{X!(N-X)!}p^X(1-p)^{N-X}.
\end{equation}
In this equation, $S$ and $B$ is the total number of counts in the source and background extraction region, $N=S+B$ and $p=1/(1+BACKSCAL)$, with $BACKSCAL$ being the area ratio of the background and source extraction regions. A smaller $P_{B}$ value indicates that a source has a larger probability of being real. For each candidate source, AE computed a $P_{B}$ value in each of the full, soft and hard bands, and we adopted the minimum of the three as the final $P_{B}$ value for the source. By setting a threshold value of $P_{B}<1\times 10^{-3}$, we obtained a final catalog of 489 X-ray sources for Terzan 5. The source positions are displayed in Figures~\ref{fig:smoothimage} and Figures~\ref{fig:roominimage}. As a comparison, we also highlight the X-ray sources detected by \citet{heinke2006} as red and green circles in the figures. Due to the much shallower exposure (i.e., $\sim 35$ ks), less than $25\%$ X-ray sources were resolved in their work. 

Because of the off-axis effect and the constraint of FoV in some of the observations, the merged images of Terzan 5 are found to have significant variation in detection sensitivity across the source searching region (Figure~\ref{fig:smoothimage}). In order to estimate the position-dependent sensitivity in the survey field, we created background and sensitivity maps following the procedures described in 47 Tuc (\citealp{cheng2019}, see also \citealp{xue2011} and \citealp{luo2017}). 
Briefly, we first masked out the detected sources from the merged event files using circular regions with mask radii ($r_{msk}$) provided by AE. We then filled in the masked region for each source with the background counts from a local annulus with inner to outer radii of $r_{msk}-2.5r_{msk}$. 
The resulted background maps were used to determine the detection sensitivity at each pixel location, which is the flux limit required for a source to be selected by our AE $P_{B}$ criterion. Given the background level at each pixel, we derived the minimum number of source counts ($S$) required for a detection using Equation (1) and the threshold $P_{B}$ value of $P_{B}=1\times 10^{-3}$. 
Due to the source crowding, the default background extraction annuli of X-ray sources strongly overlap with those neighboring sources in the cluster center. Therefore, we adopted the X-ray source parameters to estimate the detection sensitivity within $R\leq 20\arcsec$: for each source with known value of $B$ and $BACKSCAL$ given by AE, we calculated their minimum detection source counts $S$ with Equation (1) by setting $P_{B}=1\times 10^{-3}$. The obtained limiting $S$ of each source are then used to estimate the detection sensitivity profiles at the cluster central region. We found the radial profile of $S$ (i.e., the statistical minimum, median and maximum values of $S$) matches well with that of the limiting source counts map at the boundary.
Utilizing the exposure maps and assuming a simple power-law model with fixed photon-index of $\Gamma=1.4$ and column density\footnote{Such a column density is calculated from the average color excess $E(B-V)=2.38$ of Terzan 5. of $N_{H}=1.38 \times 10^{22} {\rm \ cm^{-2}}$,} we converted the limiting count rates to limiting fluxes and sensitivity maps for the main catalog in the three X-ray bands.

\subsection{Source Properties}
To derive the X-ray source properties, we utilized AE to extract the source photometry and spectra for the catalog of X-ray sources in Terzan 5. Due to the source crowding, the construction of both source apertures and background regions must take the existence of neighboring sources into consideration. These problems have been considered in AE, which was developed for the data analysis of multiple overlapping {\it Chandra} ACIS observations, especially for the fields with crowded sources \citep{broos2010}. 
Our procedures of photometry and spectra extraction for Terzan 5 are same as used in 47 Tuc \citep{cheng2019}. We built for each source a non-overlapping extraction aperture with the {\it ae\_make\_catalog} tool, and constructed the background regions with the AE ``BETTER\_BACKGROUNDS" algorithm. The default enclosed PSF power fraction of the source extraction aperture was set to be $\sim 90\%$, which was allowed to reduce to a minimum value of $\sim 40\%$ for very crowded sources. To ensure the accuracy of photometric analysis, we set a minimum number of 100 counts for each merged background spectrum during the AE workflow of ``ADJUST\_BACKSCAL" stage. The spectral extractions are first performed independently for each source and observation, then we merged the extraction results to construct composite event lists, spectra, light curves, response matrices and effective area files for each source through the ``MERGE\_OBSERVATIONS" procedure of AE. 

We computed background-subtracted photometry for each source, accounting for the energy dependent correction factors of enclosed PSF power. The aperture-corrected net source counts are derived in soft (0.5-2 keV), hard (2-8 keV) and full (0.5-8 keV) band, respectively. 
Among 489 X-ray sources, we found more than one third (i.e., $180/489$) of them have net counts greater than $\sim 30$ in the full band. We fit the spectra of these sources with the absorbed power-law and set the absorption column density as free parameters. For sources with net counts less than $\sim 30$, their spectra are poorly constrained and we converted their net count rates into unabsorbed energy fluxes using a uniform power-law model. We fixed both the photon-index ($\Gamma=1.4$) and absorption column density ($N_{H}=1.38\times 10^{22}\, \rm cm^{-2}$) during the conversion, and using the AE-generated merged spectral response files to calibrate the flux estimation. Assuming a distance of 5.9 kpc for Terzan 5, we then converted the flux of each X-ray source into unabsorbed luminosity in the soft, hard and full bands, respectively. 
Finally, we collated the source extraction and spectral fitting results into a main X-ray source catalog, sorted by the source Right Ascension. We also calculated for each source a projected distance from the cluster center, by adopting an optical central coordinate $\alpha =17^{h}48^{m}4^{s}.85$ and $\delta =-24\arcdeg 46\arcmin 44.6\arcsec$ for Terzan 5 \citep{lanzoni2010}. The final catalog of point sources is presented in Table~\ref{tab:mcat}.

\section{Analysis: Source Radial Distribution}

\begin{figure*}[htp]
\centering
\includegraphics[angle=0,origin=br,height=0.35\textheight, width=1.0\textwidth]{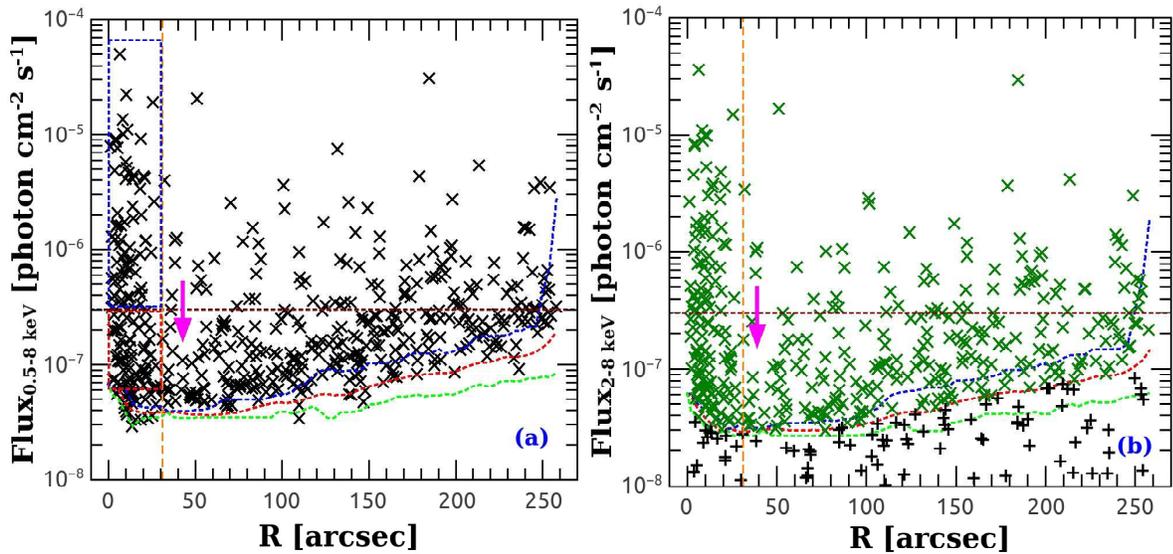}
\linespread{0.7}
\caption{Full band (a) and hard band (b) photon flux as a function of projected distance from the center of Terzan 5. The source sample selected in the hard band is shown as olive crosses in (b).
The color-coded curves represent the median (red), minimum (green) and maximum (blue) limiting detection count rate at corresponding radial bins, respectively. 
By setting a threshold photon flux of $F_{p}\sim 3.0\times 10^{-7} {\rm\ photon\ cm^{-2}\,s^{-1}}$ (horizontal wine dotted lines), the sources are divided into the faint and bright groups (see Figure-\ref{fig:surfdensefull} and Figure-\ref{fig:surfdensehard}). Samples selected by the dotted red and blue boxes in (a) are shown in Figure-\ref{fig:fit} (a). They are dynamically relaxed in Terzan 5 and have the advantage of unbiased detection.
The vertical dashed lines mark the half-mass radius ($R_h=31\arcsec$) of Terzan 5.
The magenta arrows indicate the position of the distribution dips. \label{fig:pflux}}
\end{figure*}

Due to its low Galactic latitude, Terzan 5 was found to show strong color excess (with the maximum $E(B-V)=2.82$) with significant differential extinction ($\delta E(B-V)=0.67$) over a projected angular scale of $\sim 20\arcsec$ \citep{massari2012}, which may influence our analysis of X-ray source distribution in Terzan 5. 
In the following, we study the radial distribution profiles of X-ray sources in full band (0.5-8 keV) and hard band (2-8 keV) separately. 
The total 489 detected X-ray sources are taken into consideration in the full band. We then select the sources only detected in 2-8 keV (371 sources) as the hard band sample, which may serve as a correction of foreground differential extinction for Terzan 5. 
From visual examination, we found that the surface distribution of the hard band sources is more uniform and less affected by the foreground reddening in Terzan 5.
In Figure~\ref{fig:pflux}, we plot the photon flux of the X-ray source samples as a function of projected distance from the cluster center. The full band and hard band sources are shown as black and olive crosses separately. The median, minimum and maximum value of the detection limit at each radial bin is shown as red, green and blue dotted lines, respectively. As the figure suggests, the merged images have the deepest and relatively flat limiting sensitivity near the cluster center, decreasing gradually with increasing cluster radius. 
We also found in Figure~\ref{fig:pflux} that the radial distribution of X-ray sources are sparse around $R\sim 40\arcsec$ (marked by magenta arrows). As suggested in \citet{cheng2019}, such a feature could be caused by the mass segregation effect of X-ray sources in GCs.

\begin{figure*}[htp]
\centering
\includegraphics[angle=0,origin=br,height=0.4\textheight, width=1.0\textwidth]{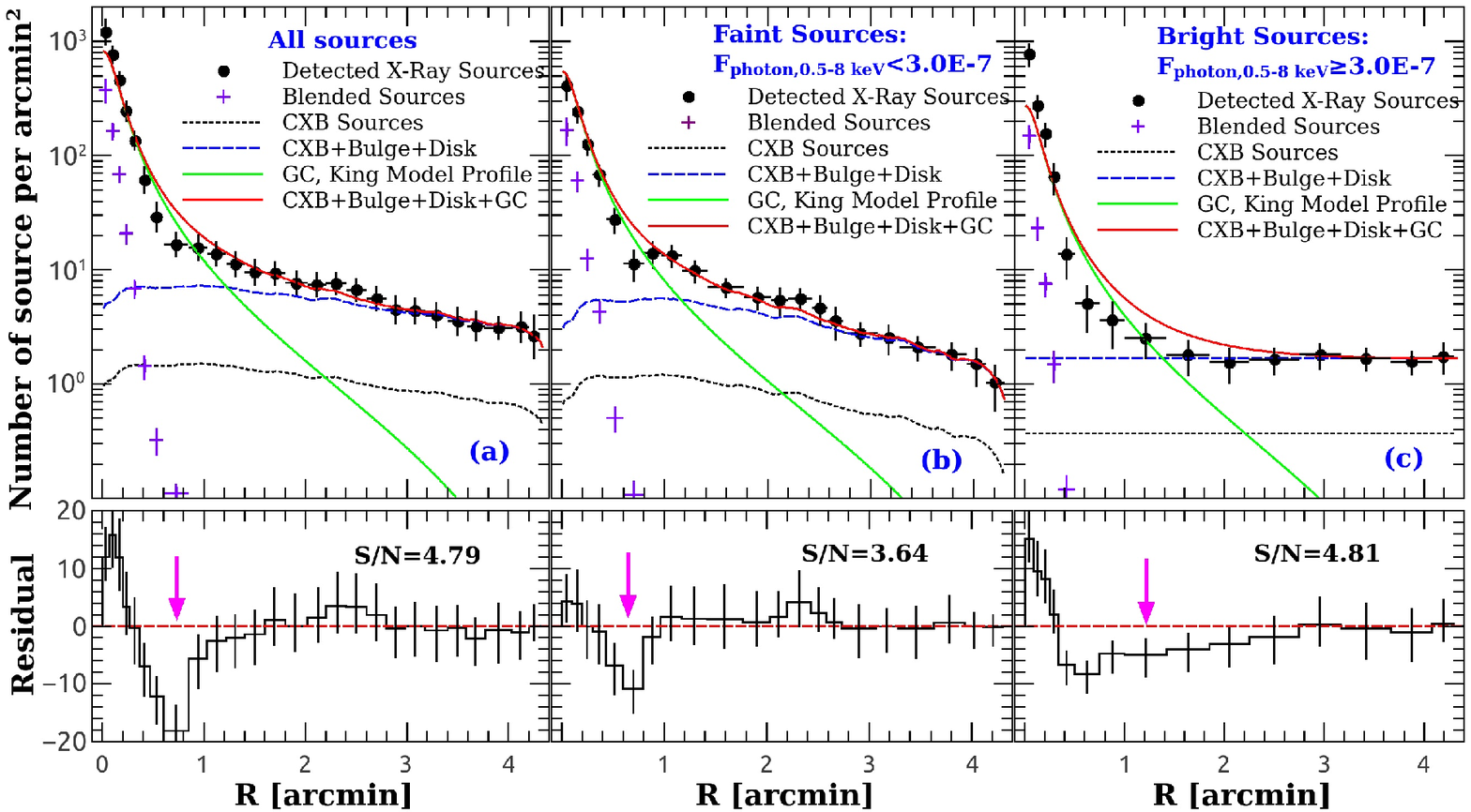}
\linespread{0.7}
\caption{Radial surface density distribution of X-ray sources in Terzan 5. The full band (0.5-8 keV) sample of X-ray sources are displayed as black dots in (a), we divided them into the faint (b) and bright (c) groups. GC sources lost by blending (i.e., $N_{B}$) in cluster center are displayed as purple pluses.
The contributions of CXB sources ($N_{CXB}$) are shown as black dotted curves, while the blue dashed curves represent the sum of CXB, Galactic Bulge and Disk X-ray sources components ($N_{G}$). The green solid lines mark the profile of stellar density convolving with the limiting sensitivity function, which was calculated from the best-fitting King model of Terzan 5 and have been normalized to match the total number of GC X-ray sources (i.e., $N_{X}+N_{B}-N_{CXB}-N_{G}$) in each group. The red solid lines represents the sum of CXB ($N_{CXB}$), Galactic Bulge and Disk ($N_{G}$), and the King model components ($N_{K}$).
The lower panels show the residual of the GC X-ray sources subtracting the King model predicted X-ray sources (i.e., $N_{X}+N_{B}-N_{CXB}-N_{G}-N_{K}$). The black text represents the maximum significance of the distribution dips.
X-ray sources are over-abundant in the cluster center and scarce at the distribution dips (marked with magenta arrows) with respect to the distribution of cluster light, and there is a small bump at $R\sim 150\arcsec$ for the faint group of X-ray sources. \label{fig:surfdensefull}}
\end{figure*}

\begin{figure*}[htp]
\centering
\includegraphics[angle=0,origin=br,height=0.4\textheight, width=1.0\textwidth]{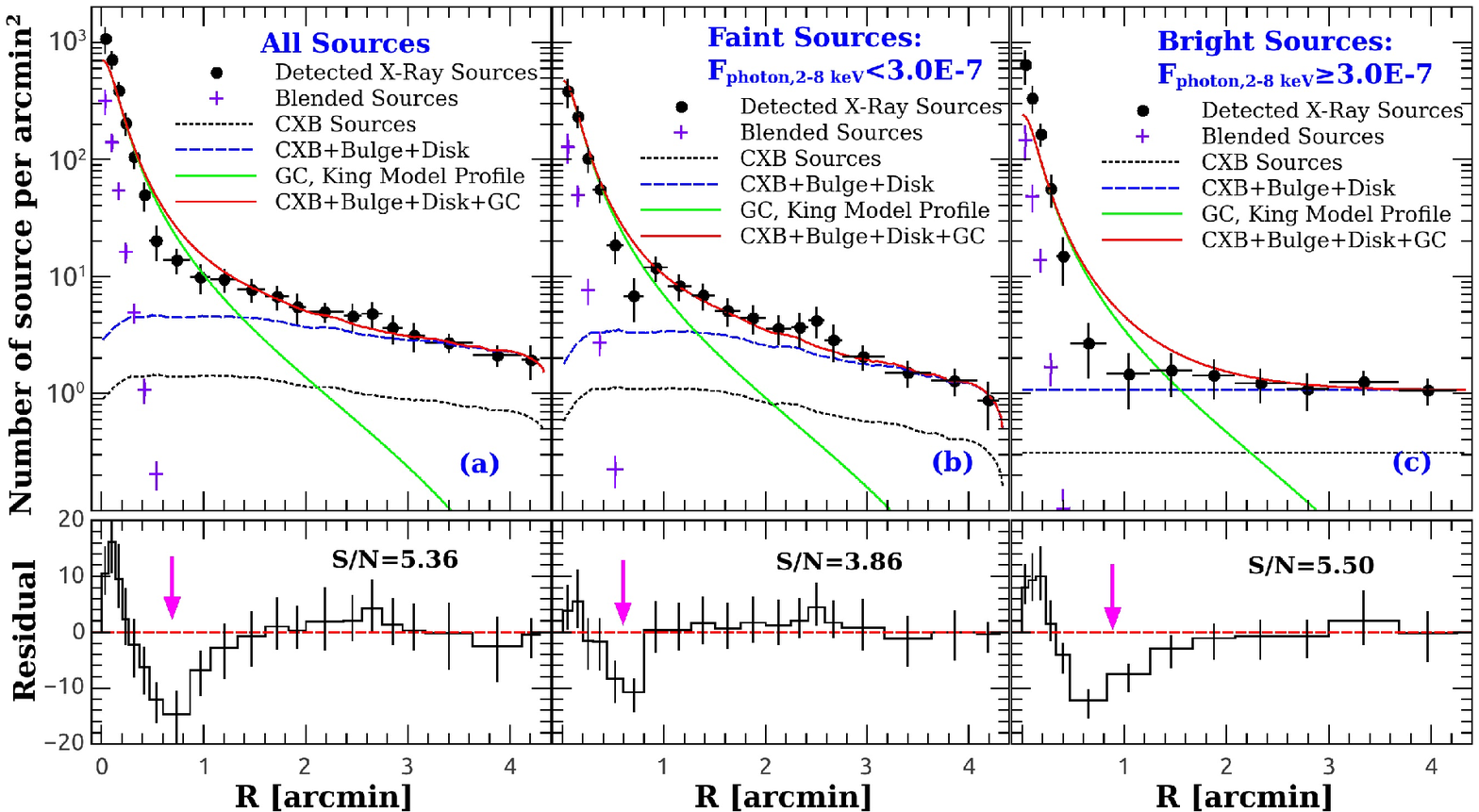}
\linespread{0.7}
\caption{Radial surface density distribution of X-ray sources in Terzan 5. Only the hard band (2-8 keV) detected source samples (i.e., olive crosses in Figure-\ref{fig:pflux}) are illustrated in this figure. Lines, texts and symbols are same as in Figure-\ref{fig:surfdensefull}. \label{fig:surfdensehard}}
\end{figure*}

To identify the possible signal of mass segregation of X-ray sources in Terzan 5, we plot the surface density of the X-ray sources as a function of projected distance from cluster center in Figure-\ref{fig:surfdensefull} and Figure-\ref{fig:surfdensehard}.
Since the mass segregation effect of X-ray sources is found to be luminosity dependent in 47 Tuc \citep{cheng2019}, we also divided the source samples into two groups according to their X-ray luminosities. 
For the full band source sample, 316 (173) of them are selected as the faint (bright) group of X-ray sources by setting a threshold photon flux of $F_{p,0.5-8}\sim 3.0\times 10^{-7} {\rm\ photon\ cm^{-2}\,s^{-1}}$ (or luminosity of $L_{X} \sim 9.5\times 10^{30} {\rm\ erg\ \,s^{-1}}$).  
While for the hard band source sample, 243 (128) are classified as the faint (bright) group of X-ray sources with a threshold photon flux of $F_{p,2-8}\sim 3.0\times 10^{-7} {\rm\ photon\ cm^{-2}\,s^{-1}}$ (or luminosity of $L_{X,0.5-8} \sim 1.6\times 10^{31} {\rm\ erg\ \,s^{-1}}$).
The surface density distributions of each group of X-ray sources are shown as black dots in the figures. The profiles are highly peaked in the cluster core, which may suggest significant populations of X-ray sources above our detection limit but were obscured by blending within the crowded central region of Terzan 5. Generally speaking, it is uncertain where sources would lie close enough together to overlap and become unresolvable, but in principle a faint source could be easily blended by brighter sources when it is located within the source extraction regions of nearby bright sources. Therefore, for each concentric annuli described by the black dots in Figure-\ref{fig:surfdensefull} and Figure-\ref{fig:surfdensehard}, we can approximately calculate the contributions of the blended sources with function 
\begin{equation}
N_{B}(>F_p)=N_{X}(>F_p)\beta.    
\end{equation} 
Here, $N_{X}(>F_{p})$ is the number of the brighter X-ray sources (i.e., sources with photon flux larger than $F_{p}$) located within the annulus, $\beta=A_{tot}/A_{ann}$, with $A_{tot}$ being the total source extraction regions of the bright X-ray sources and $A_{ann}$ donates the area of the annulus separately. The distribution profiles of $N_{B}$ are displayed as purple pluses in Figure-\ref{fig:surfdensefull} and Figure-\ref{fig:surfdensehard}. We estimated a total number of 21 (17) blended sources for the full (hard) band source sample, and about 6 (6) and 15 (11) of them are belong to the bright and faint group of X-ray sources, respectively.

With a source analyzing region much larger than the cluster core radius, our source samples are subjected to significant contamination of the cosmic X-ray background (CXB) sources, especially in the outskirts of Terzan 5. To estimate the contribution of the CXB sources, we calculated the CXB component ($N_{CXB}$) with the limiting sensitivity maps obtained in Section 3.1 and the ${\rm log}N-{\rm log}S$ relations determined by \citet{kim2007}. The cumulative number count of CXB sources above a limiting sensitivity flux $S$ can be estimated with function
\begin{equation}
{N_{CXB}(>S)}=2433(S/10^{-15})^{-0.64}-186\ {\rm deg^{-2}}
\end{equation} 
in 0.5-8 keV band, and function
\begin{equation}
{N_{CXB}(>S)}=2169(S/10^{-15})^{-0.58}-293\ {\rm deg^{-2}}
\end{equation} 
in 2-8 keV band, respectively. Here, Equation (3) and (4) are derived from the Equation (5) of \citet{kim2007}, by assuming a photon index of $\Gamma_{ph}=1.4$ and a absorption column density of $N_{H}=1.38\times 10^{22}\, \rm cm^{-2}$ for {\it Chandra} ACIS observations in the direction of Terzan 5.

The CXB components are shown as black dotted lines in Figure-\ref{fig:surfdensefull} and Figure-\ref{fig:surfdensehard}. We found the surface density profiles of observed X-ray sources significantly exceed the predicted CXB components at large cluster radius. The excess is most likely due to the population of Galactic sources ($N_{G}$), including foreground stars from the Galactic disk and possible faint X-ray sources in the Galactic bulge \citep{heinke2006}. 
Fortunately, our X-ray source searching region is much larger (with maximum radius of $R\sim260 \arcsec$, close to the tidal radius $R_t\sim 270 \arcsec$ of this cluster), and the observed stellar density profile of Terzan 5 was found to be dominant by the Galactic background stars beyond $R\sim 175\arcsec$ \citep{lanzoni2010}.
Therefore, we estimate our noncluster source numbers by looking at the radial distribution of X-ray sources outside of $R=175\arcsec$. Assuming an uniform spatial distribution for the Galactic background X-ray sources, we calculated the Galactic bulge and disk component with the limiting sensitivity maps obtained in Section 3.1, which was normalized to match the surface density profile of X-ray sources beyond $R=175\arcsec$. The sum (i.e., $N_{CXB}+N_{G}$) of the CXB and the Galactic bulge and disk components are shown as blue dashed lines in Figure-\ref{fig:surfdensefull} and Figure-\ref{fig:surfdensehard}.

From Figure-\ref{fig:surfdensefull} (a) and Figure-\ref{fig:surfdensehard} (a), the excess of X-ray sources over the blue dashed lines can be reasonably accounted for by the GC X-ray sources (i.e., $N_{X}+N_{B}-N_{CXB}-N_{G}$), which is highly peaked in cluster center and rapidly decreases with increasing $R$. Beyond $R\sim 175\arcsec$, the excess becomes insignificant and the observed X-ray sources matches well with the profile of CXB and the Galactic background components. 
As a comparison, we found that the distribution profiles of the bright sources drop more quickly than the faint sources with increasing $R$. Almost all of the bright sources are located within $R\sim 60\arcsec$ and there is no significant excess at larger cluster radii\footnote{Compared with 47 Tuc, the outer distribution peaks of the bright X-ray source could be diluted by the much higher background levels in Terzan 5. For example, the surface density of Galactic components are about $\sim 0.8-1.3\, \rm arcmin^{-2}$ in Figure-\ref{fig:surfdensefull} (c) and Figure-\ref{fig:surfdensehard} (c), which are much higher than the CXB components ($\sim 0.3-0.4\, \rm arcmin^{-2}$). While in 47 Tuc, the Galactic component is negligible, and the GC X-ray source contribution ($\sim 0.17\, \rm arcmin^{-2}$) is comparable to the CXB component ($\sim 0.26\, \rm arcmin^{-2}$; \citealp{cheng2019}) at the outskirts of the cluster. }. 
While for the faint sources, the excess over the CXB and Galactic background component is evident within $0\arcsec \lesssim R\lesssim 175\arcsec$. 
The distribution of the bright sources is more centrally concentrated than the faint sources, which may suggest a delay of sedimentation between the two groups of X-ray sources in Terzan 5.

To evaluate the significance of the mass segregation effect of X-ray sources in Terzan 5, we modeled the radial distribution of GC X-ray sources with cluster stellar density profile, by assuming that the GC X-ray sources are uniformly mixed with normal stars in Terzan 5. Following the method presented in \citet{cheng2019}, we utilized the King model \citep{king1962} to calculate the radial distribution of normal stars in Terzan 5.  
The king model parameters (i.e., the cluster concentration of $c=1.49$ and core radius of $R_{c}=9\arcsec$) are adopted from \citet{lanzoni2010}.
We convolved the King model profile with the limiting sensitivity function, and normalized the model profile to match the total number of GC X-ray sources in each group of source samples. The derived King model profiles are plotted as green solid lines in Figure-\ref{fig:surfdensefull} and Figure-\ref{fig:surfdensehard}, while the sum of CXB, King model and Galactic background components are shown as red solid lines, respectively.
Obviously, there are significant differences between the distribution profiles of observed X-ray sources and model predicted sources in Terzan 5. We calculated their residuals with equation $N_X+N_{B}-N_{CXB}-N_G-N_K$, which are illustrated as a function of $R$ in the lower panels of Figure-\ref{fig:surfdensefull} and Figure-\ref{fig:surfdensehard}.  
As shown in the figures, the residual reaches its maximum at the cluster center, decreases to the minimum at intermediate radii and rises again at larger radii, with evident distribution dips (magenta arrows) exist at $R\sim 40\arcsec$.
We defined the significance of the distribution dips as $S/N=(N_{CXB}+N_G+N_K-N_X-N_{B})/\sqrt{N_{CXB}^2\sigma_c^2+N_G^2\sigma_G^2+N_K^2\sigma_K^2+N_X^2\sigma_{P}^2(1+\beta^2)}$. Then, by adopting a nominal fitting error of $\sigma_K=\sigma_G=5\%$ for the King model fitting and Galactic background estimation, a CXB variance of $\sigma_c=22\%$ for the {\it Chandra} surveyed field within $R=4.3\arcmin$ (i.e., with an area of $\sim 0.016 \,\rm deg^{2}$) in Terzan 5 \citep{moretti2009}, and the Poisson uncertainties\footnote{Here we used the $1\sigma$ upper error of \citet{gehrels1986} to calculate the significance of the distribution dips, while both the $1\sigma$ upper and lower errors were adopted to calculate the error bars in the figures.} estimated with the formulae of \citet{gehrels1986} for the small number of X-ray sources, we adjusted the ranges of the distribution dips and estimated for each group of X-ray sources a maximum significance. 
The parameters of each distribution dip are summarized in Table-3 and the corresponding maximum significance are shown as black texts in Figure-\ref{fig:surfdensefull} and Figure-\ref{fig:surfdensehard}. 
Using numerical method, we have also simulated the distribution profiles of each group of X-ray sources with the cluster predicted models (i.e., red solid lines in Figure-\ref{fig:surfdensefull} and Figure-\ref{fig:surfdensehard}). 1000 simulations were generated separately for each group of X-ray sources, we then selected the X-ray sources within the distribution dip annuli using Poisson sampling, and simulation with selected source counts not larger than the real counts was regarded as a successful case. In Table-3, we listed the fractions of successful simulations for each group of X-ray sources, in all six groups the probabilities are very small ($<1\%$), suggesting that existence of the distribution dips are highly reliable.   

The existence of distribution dips in both full band and hard band source samples suggests that they are independent of the heavily differential extinction in the direction of Terzan 5. We also examined whether these distribution dips are created by the coverage effects (i.e., CCD gaps or variation of detection sensitivity) of the merged observations and found that such effects are negligible (Figure-\ref{fig:smoothimage}). 
Since Terzan 5 is well located within the inner region (with a projected distance of $\sim 4.3^{\circ}$ from the Galactic center) of the Galactic bulge, some of the X-ray sources are associated with the Galactic bulge and disk, and the distribution dips could be created by the enhancement of X-ray sources contribution toward the Galactic center/disk. We check this possibility by examining the azimuthal distribution of sources located within $175\arcsec \lesssim R \lesssim 258\arcsec$, and found that there is no significant excess of X-ray sources toward the Galactic center.
Taking these aspects together, we suggest that Terzan 5 is similar to 47 Tuc \citep{cheng2019}, and the radial distribution of X-ray sources in this cluster have been modified by the mass segregation effect. 
We note that the median locations and widths (i.e., ranges of $N_X+N_{B}<N_{CXB}+N_G+N_K$) of the distribution dips for the total, faint and bright groups of X-ray sources are $R_{dip}\sim 40\arcsec$, $\Delta R_{dip} \sim 70\arcsec$, $R_{dip}\sim 35\arcsec$, $\Delta R_{dip} \sim 30\arcsec$ and $R_{dip}\sim 70\arcsec$, $\Delta R_{dip} \sim 130\arcsec$, respectively.

\section{Discussion}
\subsection{Mass Segregation of X-ray Sources in Terzan 5} 
\begin{figure*}[htp]
\centering
\begin{minipage}[!htbp]{1.0\textwidth}
\leftline{\includegraphics[angle=0,origin=br,height=0.35\textheight, width=1.0\textwidth]{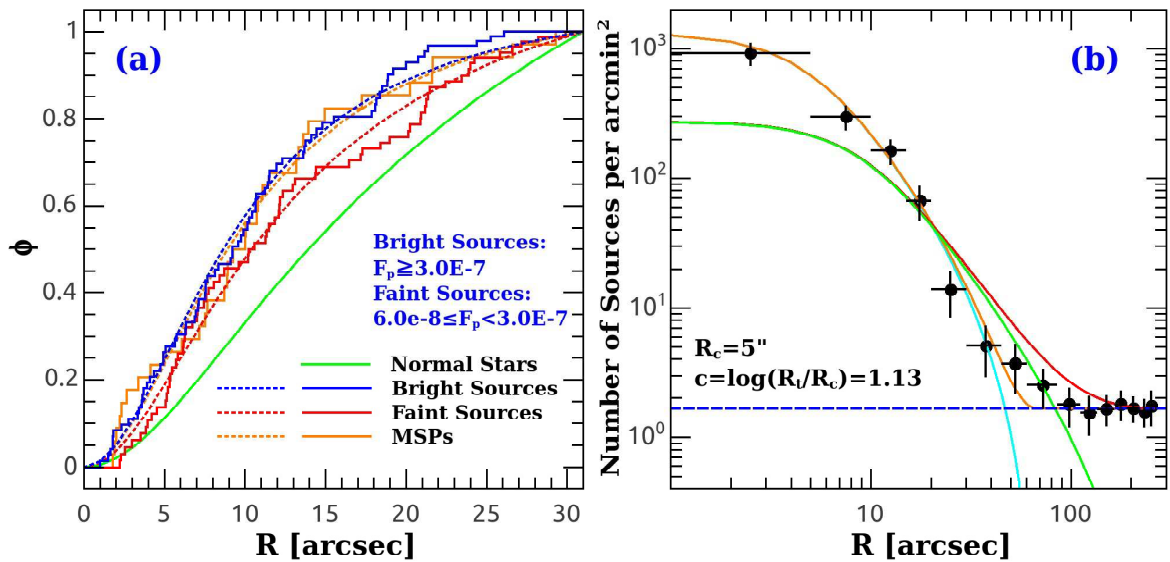}}
\linespread{0.7}
\caption{(a): Cumulative radial distribution of each group of heavy objects in Terzan 5. The orange, red and blue step lines represent the MSPs, faint ($6.0\times10^{-8}{\rm\ photon\ cm^{-2}\,s^{-1}}\lesssim F_{p}\lesssim 3.0\times 10^{-7} {\rm\ photon\ cm^{-2}\,s^{-1}}$) and bright ($F_{p}\gtrsim 3.0\times 10^{-7} {\rm\ photon\ cm^{-2}\,s^{-1}}$) group of X-ray sources, respectively. While the green solid line indicates the King model predicted distribution for normal stars. We fit the distribution of each group of heavy objects with the ``generalized King model", and the best fitting results are shown as color coded dotted lines. (b): King model fitting of the bright group of X-ray sources in Figure-\ref{fig:surfdensefull} (c). The blending corrected X-ray sources (i.e., $N_{X}+N_{B}$) are shown as black dots, and the background component (i.e., $N_{CXB}+N_{G}$) is displayed as blue dashed line. The cyan and green solid lines represent the best fitting King model of the X-ray sources ($N_{K}^{\prime}$) and the normal stars (i.e., $N_{K}$ components in Figure-\ref{fig:surfdensefull} (c)) in Terzan 5 separately, while the orange and red solid lines donate the sum of $N_{K}^{\prime}+N_{CXB}+N_{G}$ and $N_{K}+N_{CXB}+N_{G}$, respectively. \label{fig:fit}}
\end{minipage}
\end{figure*}

As described by \citet{ferraro2012}, the two-body relaxation is the main process that drives more massive object sedimentation to the cluster center, which take place with a timescale of \citep{heggie2003}
\begin{equation}
t_{relax}=0.065\frac{\sigma^{3}(R)}{G^2 M \rho(R) \rm ln \Lambda}.
\end{equation}
Here, $G$ is the gravitational constant, $\rho(R)$ and $\sigma(R)$ are the profiles of stellar density and stellar velocity dispersion, respectively, and the Coulomb logarithm $\rm ln \Lambda= ln$ $(0.11N)$, with $N$ being the total number of stars in GCs. 
Since $t_{relax}$ is anti-correlated with the mass ($M$) of the heavy object, stars with different mass are expected to drop to the cluster center with different speed.
Indeed, we have shown in Section 4 that the the locations and widths of the distribution dips for the bright X-ray sources are larger than that of the faint sources, these features are also similar to that observed in 47 Tuc, which suggests that there is a delay of sedimentation between the two groups of X-ray sources in Terzan 5.

To check the relationship of observed distribution dips with the cluster mass segregation effect, we perform the ``generalized King model" fitting to estimate the averaged mass of the faint and bright group of X-ray sources in Terzan 5. This model assumes that the X-ray sources are dynamically relaxed and they can reach a thermal equilibrium state with reference normal stars in GCs. As a result, the radial distribution of X-ray sources will be more centrally concentrated than that of the normal stars, which allows an estimation of the average mass ratio between these two groups of objects \citep{grindlay1984,grindlay2002}. 
According to the ``generalized King model", the projected surface density profile of heavy objects
takes the form (see also, \citealp{heinke2005})
\begin{equation}
S(R)=S_{0}{\biggl[1+\biggl(\frac{R}{R_{c}}\biggr)^{2}\biggr]}^{(1-3q)/2}.
\end{equation}
Here, $S_{0}$ is the normalization constant, and $q=M_{X}/M_{\ast}$ is the average mass ratio of heavy objects over the reference normal stars. $R_{c}=9\arcsec$ is the cluster core radius, which was determined using stars with masses comparable to the main-sequence turnoff stars of Terzan 5 (i.e., $M_{TO}= 0.92M_{\odot}$ for the 12 $\rm Gyr$ old and sub-solar metallicity component in Terzan 5, \citealp{lanzoni2010,ferraro2016}).

Following the steps used in 47 Tuc, we first constrain the fitting region with $R\lesssim31\arcsec$ (i.e., slightly smaller than the distribution dip of the faint group of X-ray sources) to ensure that all selected sources are dynamically relaxed in Terzan 5. Then, by setting a lower photon flux limit of $\sim 6\times 10^{-8} {\rm\ photon\ cm^{-2}\, s^{-1}}$ for the full band sources, we eliminate the possible observational bias for the source samples. 
The final sample (i.e, boxes in Figure-\ref{fig:pflux}(a)) contains 60 (68) sources for the faint (bright) group of X-ray sources, 4 (2) of them being background sources, and about 10 (6) GC sources were lost by blending. With the distribution profiles of blended sources in Figure-\ref{fig:surfdensefull}, we simulated the locations of blended sources in Terzan 5 and added them into the two source samples. Utilizing a bootstrap resamplings method proposed by \citet{grindlay2002}, we also corrected the spatial distribution of X-ray source samples for background contamination. 
In Figure-\ref{fig:fit} (a), we plot the cumulative radial distributions X-ray sources as step lines. 
Among all the considered objects, the bright X-ray sources (blue) show the highest degree of central concentration, followed by the faint X-ray sources (red) and normal reference stars (green). 
Using a maximum-likelihood method, we fit the cumulative radial distributions of X-ray sources (shown as dotted lines) with Equation (6), which gives a mass ratio of $q=1.61\pm0.12$ and $q=1.38\pm0.14$ for the bright and faint X-ray sources. Taking a main-sequence turnoff star mass of $M_{TO}= 0.92M_{\odot}$ for normal reference stars in Terzan 5, we derived an average mass of $1.48\pm0.11\,M_{\odot}$ and $1.27\pm0.13\,M_{\odot}$ for the bright and faint group of X-ray sources, respectively.
Here the errors are quoted at $1\sigma$ level.

In Figure-\ref{fig:fit} (a), we also plot the cumulative radial distribution of 34 MSPs of Terzan 5 (i.e., MSPs with $R\leqslant 31\arcsec$, \citealp{prager2017}) as orange lines.
The concentration of MSPs is comparable to the bright X-ray sources, the K-S test yields that they are different at the $3.9\%$ confidence level. While for MSPs and the faint X-ray sources, bright and the faint X-ray sources, their K-S probability of difference is $75.7\%$ and $93.9\%$, respectively. 
We fit the cumulative radial distribution of MSPs with the ``generalized King model", the maximum-likelihood method gives a best-fitting mass ratio of $q=1.57\pm0.15$ for the MSPs, which agrees well with the value (i.e., $q=1.61\pm0.12$) of the bright X-ray sources. For comparison, our mass ratio of ``generalized King model" fitting for MSPs and bright X-ray sources is consistent with the result of \citet{prager2017}. By using a  nonlinear least-squares fit for Equation (6), \citet{prager2017} obtained an average mass ratio of $q=1.57\pm0.19$ for MSPs and LMXBs in Terzan 5. 
Considering that both MSPs and LMXBs are dynamically formed in GCs, these findings may suggest that many of the bright X-ray sources in Terzan 5 have a dynamical origin. 
As discussed in 47 Tuc \citep{cheng2019}, the mass segregation is effectual in driving primordial binaries sedimentation into the dense core of GCs, which enhances the encounter probability of binaries and contributes to the population of bright X-ray sources in GCs.
We note that the estimated average mass for the bright and faint group of X-ray sources in Terzan 5 is slightly larger that found in 47 Tuc, with $1.48\pm0.11\,M_{\odot}$ and $1.27\pm0.13\,M_{\odot}$ in Terzan 5, and $1.44\pm0.15\,M_{\odot}$ and $1.16\pm0.06\,M_{\odot}$ in 47 Tuc, respectively.
These discrepancies may result from the higher threshold of X-ray luminosity for the bright and faint groups of X-ray sources in Terzan 5 (i.e., $L_{X} \sim 9.5\times 10^{30} {\rm\ erg\ \,s^{-1}}$) than in 47 Tuc (i.e., $L_{X} \sim 5.0\times 10^{30} {\rm\ erg\ \,s^{-1}}$).

Our estimation of average mass for the bright and faint groups of X-ray sources is consistent with the observed distribution dips in Section 4. Namely, with a larger average mass than the faint X-ray sources, the bright X-ray sources are expected to drop to the cluster center faster under the effect of two-body relaxation, and their distribution dip will propagate outward further in Terzan 5. 
We note that the sedimentation delay of X-ray sources in Terzan 5 is consistent with that found in 47 Tuc \citep{cheng2019}, suggesting a universal mass segregation effect for X-ray sources in GCs.

Theoretically, with the sedimentation and concentration of heavy objects in cluster center, the average mass of stars (and $t_{relax}$) in cluster core would increases (decreases) significantly. As a consequence, the cluster core will evolve more and more faster than the outer parts of GCs, and mass segregation may lead to the formation of a new cluster consist of heavy objects in the center of GCs \citep{spitzer1987}. In Section 4, it can be seen that the distribution profiles of the faint X-ray sources are roughly consistent with that of normal stars in Terzan 5, while for the bright X-ray sources, the discrepancy between these two profiles is very large. In fact, we found that the distribution profile of the bright X-ray sources can be better fitted with a new King model. As shown in Figure-\ref{fig:fit} (b), the best-fitting King model (cyan and orange solid lines) gives a core radius of $R_{c}=5\arcsec$ and cluster concentration of $c=1.13$ for the bright X-ray sources\footnote{Here we only displayed the results of the full band source sample, since the fitting results are same for the hard band source sample.}, which is more dense and more compact than the distribution of the normal stars (i.e., green and red solid lines, with $R_{c}=9\arcsec$ and $c=1.49$, respectively), indicating that the bright X-ray sources are successfully segregated from normal stars in Terzan 5. 

\subsection{Dynamical Evolution of GCs in Galactic Field: Terzan 5 versus 47 Tuc}
With a total mass of $M_c \simeq 2 \times 10^{6}M_{\odot}$ \citep{lanzoni2010}, Terzan 5 is one of the most massive clusters in the Milky Way \citep{harris1996}. 
However, the finding of two significantly different stellar populations of Terzan 5 suggests that the star formation process in the cluster was not finished in one episode, but characterized by distinct bursts. 
After an initial period of star formation, which happened $\sim 12\, \rm Gyr$ ago and is responsible for the creation of the sub-solar components of the system, Terzan 5 was thought to experience
a long phase ($\sim 7.5 \, \rm Gyr$) of quiescence. Then, approximately $4.5 \, \rm Gyr$ ago, another star formation burst occurred and generated a super-solar population of stars in the cluster \citep{ferraro2009,ferraro2016}. 
There are many models to explain these unique characteristics of Terzan 5, which include the nucleus of a defunct dwarf galaxy accreted by the Milky Way (\citealp{ferraro2009}; a possible case for this scenario is $\omega$ Centauri; \citealp{bekki2003}), a remnant of a pristine fragment or `fossil remnant' of the Galactic bulge \citep{lanzoni2014,massari2014,ferraro2016}, a massive star cluster underwent `self-enrichment' in order to induce an additional epoch of star formation\footnote{It is worth to point out that the `self-enrichment' scenario is facing serious problems, since the fraction of chemical enriched stars was found to be dominated in most GCs \citep{bastian2015,milone2017,bastian2018,reina2018}.} \citep{ferraro2016}, or a collision of GC with a giant molecular cloud and lead to the generation of a super-solar population of stars \citep{mckenzie2018}.
All of these models require that the original mass of Terzan 5 was much larger ($M_c \gtrsim 10^{7-8}M_{\odot}$) in the past than what is observed today, so that it can retain or capture a large amount of metal-enriched gas and trigger another star formation process in the system. 

Besides the peculiar stellar populations and huge mass, the structure of Terzan 5 was found to be very compact. 
The core, half-mass and tidal radius of Terzan 5 is $R_{c}\simeq 0.26\,\rm pc$, $R_{h}\simeq 0.89\,\rm pc$ and $R_{t}\simeq 7.96\,\rm pc$, respectively \citep{lanzoni2010}. 
As a comparison, we found that the total mass of 47 Tuc ($M_c \sim 1 \times 10^{6}M_{\odot}$; \citealp{cheng2018a}) is slightly lower than Terzan 5, but its size is much larger, with $R_{c}\simeq 0.41\,\rm pc$, $R_{h}\simeq 3.71\,\rm pc$ and $R_{t}\simeq 41.56\,\rm pc$, respectively \citep{mcLaughlin2006}.
These evidences indicate that the stellar density in Terzan 5 is much larger than 47 Tuc, and accordingly the stellar dynamical processes in Terzan 5 will take place more frequently. 
In fact, the central stellar density of Terzan 5 was estimated to be $\rho_c\simeq 1.58\times 10^{6}\, \rm M_{\odot}\, pc^{-3}$ \citep{prager2017}, which is at least an order of magnitude greater than 47 Tuc ($\rho_c\simeq 1.19\times 10^{5}\, \rm M_{\odot}\, pc^{-3}$; \citealp{prager2017}).
Considering that the stellar velocity dispersion in Terzan 5 and 47 Tuc is $\sigma \sim 18\, \rm km\,s^{-1}$ and $\sigma \sim 11.5\,\rm km\,s^{-1}$ \citep{baumgardt2018}, respectively, and the age of Terzan 5 ($\tau\sim 12\,\rm Gyr$ or $\tau\sim 4.5\,\rm Gyr$; \citealp{ferraro2016}) is comparable to 47 Tuc ($\tau=13.06\,\rm Gyr$; \citealp{forbes2010}), according to Equation (5), it is reasonable to infer that the dynamical age (i.e., $t_{d}\simeq \tau/t_{relax}$) of Terzan 5 is older than 47 Tuc.

\begin{figure}[htp]
\leftline{\includegraphics[angle=0,origin=br,height=0.33\textheight, width=0.5\textwidth]{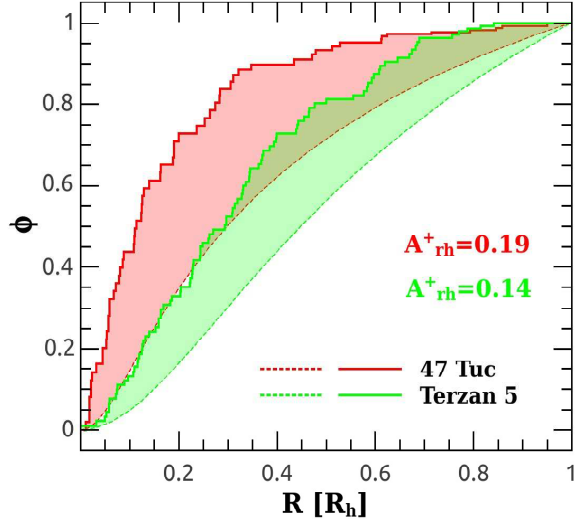}}
\linespread{0.7}
\caption{Cumulative radial distribution of X-ray sources (with unabsorbed luminosity of $L_{X}\gtrsim 5.0\times 10^{30}{\rm\ erg\,s^{-1}}$) and reference stars (i.e, the King model distribution of normal stars) in Terzan 5 (green lines) and 47 Tuc (red lines). The area of the shaded regions ($A^{+}_{rh}$) is calculated with Equation (7), which is a good indicator of cluster dynamical age. \label{fig:age}}
\end{figure}

On the other side, with the observed distribution profiles of X-ray sources in Terzan 5 and 47 Tuc, we can assess the dynamical age of these two clusters with the mass segregation effect.
As suggested by \citet{ferraro2012}, the projected radius (in units of $R_c$) of the distribution dip is a good indicator of the dynamical age in GCs (i.e., $t_d \propto 1/t_{relax} \propto R_{dip}$).
For Terzan 5, we found $R_{dip}\simeq 70\arcsec\simeq 7.8\,R_c$ for the bright sources and $R_{dip}\simeq 35\arcsec\simeq 3.9\,R_c$ for faint sources in Figure-\ref{fig:surfdensefull}. While for 47 Tuc, the corresponding values are $R_{dip}\simeq 170\arcsec \simeq 8.2\,R_c$ and $R_{dip}\simeq 100\arcsec \simeq 4.8\,R_c$ \citep{cheng2019}, respectively. At first sight, $R_{dip}$ in 47 Tuc is slightly larger than that of Terzan 5, which is in conflict with what $t_{relax}$ suggests. 
However, the precision of $R_{dip}$ is affected by the binning algorithm of the heavy objects, which may hamper the comparison of dynamical age between these two GCs. 
By using direct N-body simulations, \citet{alessandrini2016} proposed a binning-independent parameter ($A^{+}$) to characterize the progressive mass segregation of heavy objects in GCs, which is found to be more precise than $R_{dip}$ and can be applied to all GCs \citep{lanzoni2016,ferraro2018}. $A^{+}$ is defined as the area enclosed between the cumulative radial distribution of heavy objects (here the X-ray sources $\phi_{X}(R)$) and that of a reference stars ($\phi_{REF}(R)$), with
\begin{equation}
A^{+}(R)=\int_{R_0}^{R}\phi_X(R)-\phi_{REF}(R)dR,
\end{equation}
where, $R_0$ and $R$ are the innermost and the outermost distances from the cluster center considered in the analysis.

In Figure-\ref{fig:age}, we assess the dynamical age of Terzan 5 and 47 Tuc with parameter $A^{+}$. Following \citet{lanzoni2016} and \citet{ferraro2018}, we measure $A^{+}$ from the cluster center to the half-light radius (hereafter,$A^{+}_{rh}$). Since the mass segregation of X-ray sources is luminosity dependent, we first set a uniform threshold unabsorbed luminosity of $L_{X}\simeq 5.0\times 10^{30}{\rm\ erg\,s^{-1}}$ for both clusters.  
Then, we corrected the source blending and CXB contamination for the source samples, and their cumulative radial distributions are plotted as solid step lines in Figure-\ref{fig:age}.
For the reference stars, which are calculated from the King model profiles of these two clusters and are shown as dotted lines in the figure. 
Finally, we obtained a value of $A^{+}_{rh}=0.14$ for Terzan 5 and $A^{+}_{rh}=0.19$ for 47 Tuc, respectively. The larger value of $A^{+}_{rh}$ suggests that the cluster is dynamically older \citep{ferraro2018}. Again, these results are in conflict with the values of $t_{relax}$ in Terzan 5 and 47 Tuc. 

Considering that the orbit of Terzan 5 is short and with high eccentricity (i.e., $R_{p}=0.82\pm0.32\, \rm kpc$ and $R_{a}=2.83\pm 0.54\, \rm kpc$) in the Galactic field\footnote{As a comparison, the perigalactic and apogalactic distance of 47 Tuc is $R_{per}=5.46\pm 0.01 \,\rm kpc$ and $R_{apo}=7.44\pm 0.02 \,\rm kpc$, respectively \citep{baumgardt2019}.}, its current mass is very massive and the stellar populations suggest that its original mass was even more massive in the past, the stripping of tidal shock may provide a reasonable explanation for the discrepancy in dynamical ages between Terzan 5 and 47 Tuc.
Like all self-gravitating systems, the evolution of GCs is balanced by the production of energy in the core and the outflow of energy from the cluster. Although the central energy source plays an important role in heating the system, but the energy production rate of GCs is controlled by the structure of the system as a whole \citep{henon1961}. 
For GCs in isolation, the rate of enegy-outflow is mainly driven by two-body relaxation, which lead to the escape of low-mass stars from the cluster and result in the expansion of GCs \citep{henon1965,gieles2011}. 
For comparison, GCs in tidal field will suffer enhanced rate of energy-outflow when stars are evaporating outside of the cluster Jacobi radius by tidal force. 
In response, the cluster will contract homologically and the increase in stellar density is helpful to enhance the energy production rate of the cluster core \citep{gieles2011}. 
Therefore, the structure of GCs in tidal field (or close to Galactic center) will become more compacted than that of isolated GCs (or GC far away from Galactic center), and tidal stripping may plays an important role in accelerating the dynamical evolution of GCs \citep{gnedin1999}.

Furthermore, the enhanced energy production rate of Terzan 5 suggests that the ``binary burning" process will take place more frequently in the cluster core. As a consequence, more exotic objects are expected to be created in this cluster.
This is in agreement with the finding of a large number of LMXBs \citep{bahramian2014} and MSPs \citep{ransom2005,prager2017} in Terzan 5. 
In fact, both the stellar encounter rate (i.e., $\Gamma$; \citealp{bahramian2013}) and specific stellar encounter rate (i.e., $\gamma=\Gamma/M_c$; \citealp{pooley2006,cheng2018a}) of Terzan 5 are found to be the largest among Galactic GCs, which may responsible for the creation of a very peculiar LXMB, IGR J1748--2446, in Terzan 5 \citep{strohmayer2010,papitto2011}, since formation of this system may suffer from more than one close dynamical encounters \citep{jiang2013}.

\section{Summary}
We have presented a sensitive study of weak X-ray sources in Terzan 5. The main points of this work are summarized below.

1. We detected 489 X-ray sources within a total area of $58\arcmin .1^{2}$ (or within a radius of $4\arcmin .3$) in Terzan 5. The cleaned net exposure time of the study is 734 ks and more than $75\%$ of the sources are first detected in this cluster.
 
2. We show that there are significant distribution dips in the radial distribution profiles of weak X-ray sources in Terzan 5, the location and width of the distribution dip for the total, faint ($L_{X} \lesssim 9.5\times 10^{30} {\rm\ erg\ \,s^{-1}}$) and bright ($L_{X} \gtrsim 9.5\times 10^{30} {\rm\ erg\ \,s^{-1}}$) source samples is $R\sim 40\arcsec$, $\Delta R \sim 70\arcsec$, $R\sim 35\arcsec$, $\Delta R \sim 30\arcsec$ and $R\sim 70\arcsec$, $\Delta R \sim 130\arcsec$, respectively. And there is a delay of sedimentation for the faint and bright groups of X-ray sources in Terzan 5.

3. By fitting the radial distribution of the bright and faint groups of X-ray sources with a ``generalized King model", we estimated an average mass of $1.48\pm0.11\,M_{\odot}$ and $1.27\pm0.13\,M_{\odot}$ for the bright and faint group of X-ray sources separately. These results are qualitatively consistent with the observed distribution dips in Terzan 5, and suggest that mass segregation may be responsible for the creation of these distribution features.

4. The observed sedimentation delay of X-ray sources in Terzan 5 is in agreement with that found in 47 Tuc \citep{cheng2019}, which may indicate a universal mass segregation effect for X-ray sources in GCs.

5. Compared with 47 Tuc, we found that the stellar density (two-body relaxation timescale) profile in Terzan 5 is much larger (smaller), but its dynamical age is younger than 47 Tuc. These features suggest that the evolution of Terzan 5 is not purely driven by two-body relaxation, and tidal stripping effect may play an important role in accelerating the dynamical evolution of this cluster. 

\begin{deluxetable}{lrrcrrrcrrrc}
\tabletypesize{\scriptsize}
\tablecolumns{12}
\linespread{1}
\tablewidth{0pc}
\tablenum{1}
\tablecaption{Log of {\it Chandra} Observations}
\tablehead{
\colhead{ObsID} & \colhead{Date} & \colhead{Arrays} & \colhead{Livetime} & \colhead{Ra} & \colhead{Dec} & \colhead{Roll} & \colhead{Frame}  & \colhead{$\delta x$} & \colhead{$\delta y$} & \colhead{Rotation} & \colhead{Scale}\\
\colhead{} & \colhead{} & \colhead{} & \colhead{(ks)} & \colhead{(\arcdeg)} & \colhead{(\arcdeg)} & \colhead{(\arcdeg)} & \colhead{time (s)} & \colhead{(pixel)} & \colhead{(pixel)} & \colhead{(\arcdeg)} & \colhead{Factor}\\
\colhead{(1)} & \colhead{(2)} & \colhead{(3)} & \colhead{(4)} & \colhead{(4)} & \colhead{(6)} & \colhead{(7)} & \colhead{(8)}  & \colhead{(9)} & \colhead{(10)} & \colhead{(11)} & \colhead{(12)}}
\startdata
3798  & 2003-07-13 & ACIS-S & 39.34  & 267.0202 & -24.7899 & 272.19& 3.04 & -0.412 & -1.328 & -0.014 & 1.0006 \\
10059 & 2009-07-15 & ACIS-S & 36.26  & 267.0238 & -24.7840 & 271.84& 1.64 & -0.477 & -1.289 & -0.047 & 1.0002 \\
13225 & 2011-02-17 & ACIS-S & 29.67  & 267.0172 & -24.7759 & 89.69 & 3.04 & -0.363 & -1.415 &  0.007 & 1.0004 \\
13252 & 2011-04-29 & ACIS-S & 39.54  & 267.0174 & -24.7758 & 87.69 & 3.04 & -0.193 & -0.498 & -0.022 & 0.9998 \\
13705 & 2011-09-05 & ACIS-S & 13.87  & 267.0245 & -24.7815 & 269.31& 3.04 &  0.024 & -0.612 & -0.014 & 1.0008 \\
13706 & 2012-05-13 & ACIS-S & 46.46  & 267.0190 & -24.7781 & 86.89 & 3.04 & -0.812 & -1.718 &  0.028 & 1.0013 \\
14339 & 2011-09-08 & ACIS-S & 34.06  & 267.0245 & -24.7815 & 269.31& 3.04 & -0.065 & -0.667 &  0.100 & 1.0011 \\ 
14475 & 2012-09-17 & ACIS-S & 30.5   & 267.0253 & -24.7788 & 269.04& 0.84 & -0.330 & -0.983 &  0.023 & 0.9995 \\
14476 & 2012-10-28 & ACIS-S & 28.6   & 267.0253 & -24.7789 & 267.99& 0.84 & -0.157 & -0.176 & -0.040 & 1.0007 \\
14477 & 2013-02-05 & ACIS-S & 28.6   & 267.0197 & -24.7755 & 90.14 & 0.84 & -0.769 & -1.013 & -0.011 & 0.9984 \\
14478 & 2013-07-16 & ACIS-S & 28.6   & 267.0251 & -24.7791 & 263.16& 0.84 & -0.745 & -1.191 &  0.138 & 0.9987 \\
14479 & 2014-07-15 & ACIS-S & 28.6   & 267.0255 & -24.7788 & 271.91& 0.84 & -0.831 & -1.290 & -0.003 & 1.0014 \\
14625 & 2013-02-22 & ACIS-S & 49.2   & 267.0189 & -24.7781 & 89.16 & 3.04 & -0.126 & -0.333 & -0.006 & 1.0011 \\
15615 & 2013-02-23 & ACIS-S & 84.16  & 267.0189 & -24.7782 & 89.16 & 3.04 &  0.000 &  0.000 &  0.000 & 1.0000 \\
15750 & 2014-07-20 & ACIS-S & 22.99  & 267.0247 & -24.7813 & 275.16& 3.04 & -0.888 & -1.538 & -0.063 & 1.0006 \\
16638 & 2014-07-17 & ACIS-S & 71.6   & 267.0247 & -24.7813 & 275.16& 3.04 & -0.435 & -0.374 &  0.074 & 1.0008 \\
17779 & 2016-07-13 & ACIS-S & 68.85  & 267.0244 & -24.7803 & 266.16& 3.04 &  0.673 &  0.273 &  0.016 & 0.9998 \\
18881 & 2016-07-15 & ACIS-S & 64.71  & 267.0244 & -24.7803 & 266.16& 3.04 &  0.643 &  0.174 &  0.176 & 1.0003 \\
\enddata
\vspace{-0.5cm}
\tablecomments{Columns.\ (1) -- (3): {\it Chandra} observation ID, date and instrument layout for Teran 5. Columns.\ (4) -- (8): livetime, telescope optical pointing position, roll angle and frame time of each observation. Columns.\ (9) -- (12): right ascension and decl. shift, rotation angle and scale factor of {\it wcs\_match} transformation matrix.}
\label{tab:obslog}
\end{deluxetable}

\begin{deluxetable}{cllccccccccc}
\tabletypesize{\scriptsize}
\tablewidth{0pt}
\tablecaption{Main {\it Chandra} Source Catalog}
\tablecolumns{12}
\linespread{1}
\tablewidth{0pc}
\tablenum{2}
\tablehead{
\colhead{XID} & \colhead{RA} & \colhead{Dec} & \colhead{Error} & \colhead{R} & \colhead{$\log P_{\rm B}$} & \colhead{$C_{net,f}$} & \colhead{$C_{net,h}$} & \colhead{$F_{p,f}$} & \colhead{$F_{p,h}$} & \colhead{$\log L_{X,f}$} & \colhead{$\log L_{X,h}$} \\
\colhead{} & \colhead{(\arcdeg)} & \colhead{(\arcdeg)} & \colhead{(\arcsec)} & \colhead{(\arcsec)} & \colhead{} & \colhead{(cts)}  & \colhead{(cts)}  & \colhead{(ph/s/cm$^{2}$)} & \colhead{(ph/s/cm$^{2}$)} & \colhead{(erg/s)} & \colhead{(erg/s)} \\
\colhead{(1)} & \colhead{(2)} & \colhead{(3)} & \colhead{(4)} &\colhead{(5)} & \colhead{(6)} & \colhead{(7)} & \colhead{(8)} & \colhead{(9)} & \colhead{(10)} & \colhead{(11)} & \colhead{(12)}
}
\startdata
1 & 266.947749 & -24.762933 & 0.2 & 243.9 & $<$-5 & $63.2^{+9.1}_{-9.1}$ & $45.4^{+7.8}_{-7.4}$ & 6.44E-7 & 4.99E-7 & 31.29 & 31.12 \\ 
2 & 266.949969 & -24.759574 & 0.1 & 240.1 & $<$-5 & $186^{+14}_{-14}$    & $143^{+12}_{-12}$    & 1.56E-6 & 1.29E-6 & 31.74 & 31.60 \\ 
3 & 266.950132 & -24.800169 & 0.2 & 241.3 & $<$-5 & $28.6^{+7.3}_{-6.6}$ & $20.3^{+6.1}_{-5.5}$ & 2.38E-7 & 1.82E-7 & 30.86 & 30.70 \\ 
4 & 266.950526 & -24.779173 & 0.2 & 227.8 & $<$-5 & $30.4^{+7.3}_{-6.8}$ & $13.8^{+5.5}_{-4.9}$ & 2.41E-7 & 1.18E-7 & 30.87 & 30.71 \\ 
5 & 266.955418 & -24.772218 & 0.0 & 213.2 & $<$-5 & $693^{+27}_{-27}$    & $486^{+23}_{-22}$    & 5.44E-6 & 4.16E-6 & 32.23 & 32.05 \\ 
6 & 266.956670 & -24.777386 & 0.1 & 207.8 & $<$-5 & $61.4^{+8.9}_{-8.7}$ & $62.2^{+8.5}_{-8.5}$ & 4.74E-7 & 5.26E-7 & 31.49 & 31.40 \\ 
7 & 266.957447 & -24.767785 & 0.2 & 209.1 & $<$-5 & $25.3^{+6.6}_{-6}$   & $5.7^{+4}_{-3.4}$    & 1.93E-7 & 4.81E-8 & 30.76 & 30.61 \\ 
8 & 266.957524 & -24.774669 & 0.1 & 205.5 & $<$-5 & $103^{+11}_{-11}$    & $52.2^{+7.9}_{-7.9}$ & 7.91E-7 & 4.42E-7 & 31.53 & 31.06 \\ 
\enddata
\vspace{-0.5cm}
\tablecomments{
Column\ (1): sequence number of the X-ray source catalog, sorted by R.A.
Columns.\ (2) and (3): right ascension and decl. for epoch J2000.0.
Columns.\ (4) and (5): estimated standard deviation of the source position error and its projected distance from cluster center.
Column.\ (6): logarithmic Poisson probability of a detection not being a source.
Columns.\ (7) -- (10): net source counts and photon flux extracted in the full (0.5--8~keV) and hard (2-8~keV) band, respectively.
Columns.\ (11) and (12): unabsorbed source luminosity in full and hard band. 
The full content of this table is available online.
}
\label{tab:mcat}
\end{deluxetable}

\begin{deluxetable}{@{}ll*{9}{c}rrr@{}}
\tabletypesize{\small} 
\tablecolumns{9}
\linespread{1}
\tablewidth{0pc}
\tablenum{3}

\tablecaption{Significance of the Distribution Dips}
\tablehead{
\colhead{Source Groups} & \colhead{R} & \colhead{$N_{\rm X}$} & \colhead{$N_{\rm B}$} & \colhead{$N_{\rm CXB}$} & \colhead{$N_{\rm G}$} & \colhead{$N_{\rm K}$} & \colhead{$S/N$} & \colhead{Prob.}\\
\colhead{(1)} & \colhead{(2)} & \colhead{(3)} & \colhead{(4)} & \colhead{(5)} & \colhead{(6)} & \colhead{(7)} & \colhead{(8)} & \colhead{(9)}}
\startdata
\multicolumn{9}{c}{Full Band ($0.5-8$ keV) Source Samples}\\
\hline
All Sources    & 39-46   & 3  & 0.06 & 0.76 & 2.91 & 13.90 & 4.79 & 0/1000 \\
Faint Sources  & 36-46   & 5  & 0.09 & 0.83 & 3.03 & 13.89 & 3.64 & 7/1000\\
Bright Sources & 22-77   & 16 & 0.05 & 1.75 & 6.21 & 34.04 & 4.81 & 0/1000 \\
\hline
\multicolumn{9}{c}{Hard Band ($2-8$ keV) Source Samples} \\                              
\hline
All Sources    & 27-50   & 20 & 0.19 & 2.17 & 4.82 & 45.54 & 5.36 & 0/1000 \\
Faint Sources  & 25-50   & 18 & 0.16 & 1.79 & 3.70 & 34.33 & 3.86 & 4/1000 \\
Bright Sources & 22-77   & 10 & 0.05 & 1.48 & 3.60 & 30.01 & 5.50 & 0/1000 \\
\enddata
\vspace{-0.5cm}
\tablecomments{
Col.\ (1): name of the source groups defined in Figure-\ref{fig:surfdensefull} and Figure-\ref{fig:surfdensehard}.
Cols.\ (2): the annulus ranges (in units of arcsec) of the distribution dip when its significance  reach a maximum value.
Col.\ (3)-Col.\ (7): number of detected X-ray sources, possible count of blended GC sources, CXB sources, Galactic background sources and the King model predicted sources within the annulus region.
Col.\ (8): the maximum significance of the distribution dip. 
Col.\ (9): fraction of the 1000 simulations show a distribution dip with strength similar to what is observed (i.e., with Poisson resampled sources count $N\leq N_X$ within the annulus region).}
\label{tbl:spec_freq}
\end{deluxetable}

\begin{acknowledgements}
We thank the anonymous referee for the valuable suggestions that helped to improve our manuscript. This work is supported by the National Key R\&D Program of China No. 2017YFA0402600, No. 2017YFA0402703, and No. 2016YFA0400803, the National Natural Science Foundation of China under grants 11525312, 11890692, 11133001, 11333004, 11773015, 11303015, and 11873029, and the Project U1838201 supported by NSFC and CAS.

\end{acknowledgements}

\label{lastpage}

\end{document}